\documentstyle[pra,preprint,aps,subfigure]{revtex}

\begin{document}
     
\draft
\title{Semiclassical Dynamics of the Jaynes-Cummings Model}

\author{Adrian Alscher and Hermann Grabert}

\address{Fakult\"at f\"ur Physik, Albert-Ludwigs-Universit\"at,\\
Hermann-Herder-Strasse 3, D-79104 Freiburg, Deutschland}
\date{\today}
\maketitle 
\tighten
   
\begin{abstract}
The  semiclassical approximation of coherent state path integrals is
employed to study  the dynamics of the Jaynes-Cummings
model. Decomposing the Hilbert space into subspaces of given
excitation quanta above the ground state, the  semiclassical
propagator is shown to describe the exact quantum dynamics of the
model. 
We also present a semiclassical approximation that does not exploit
the special properties of the Jaynes-Cummings Hamiltonian and can be
extended to more general situations. In this approach the  
contribution of the dominant
semiclassical paths and the relevant fluctuations about them are
evaluated. This theory  leads to an accurate description of  
spontaneous emission going beyond the usual classical field 
approximation.  
\end{abstract}


\narrowtext

\section{Introduction}
Since the sixties of the last century the Jaynes-Cummings model
\cite{jaynes} is frequently considered as a simple model 
to describe  a two-level atomic system interacting with an
electromagnetic field in a cavity; for recent reviews see 
 \cite{stenholm,shoreknight}. 
Apart from its relevance to quantum optics, in particular laser
theory,  this integrable quantum model also allows to test
approximative methods by comparing them with the exact result.  
In particular, the ``semiclassical'' theory has attracted considerable
attention where the bosonic field mode is represented by classical
c-numbers while the two-level atomic system is represented as a
quantum  spin-$\frac{1}{2}$ \cite{crisp}-\cite{foerster}. 
In this approximation the Heisenberg equations of motion
\cite{ackerhalt} are replaced by linear operator equations for the
spin variables and an amplitude equation for the electromagnetic  field
which is driven by the expectation values of the spin operators. Taking the
expectation value of the  Heisenberg equations for the spin variables, the
optical Bloch equations emerge which  describe the dynamics of a
classical Bloch vector on the two-sphere \cite{louiselle,allen}. 
It is well known that this ``semiclassical'' theory provides results that
are equivalent to a full quantum mechanical treatment if the mean
number of bosons is very large and fluctuations in the boson number
can be neglected \cite{gea}.

While this conventional semiclassical approach treats the  cavity
field just classically, we attempt at a semiclassical theory treating
both the atomic and electromagnetic subsystems on an equal footing.
Starting from the full quantum model we focus on the most probable
paths of the system within the path 
integral representation and relevant fluctuations about them. Within
the scope  of (spin) coherent  state path integrals we obtain 
a semiclassical approximation going beyond the classical field
approximation. For instance, the approach  yields an accurate
description of spontaneous emission.

The paper is organized as follows. In Sec.II we first solve the
Jaynes-Cummings model exactly with spin coherent state path integrals  
in a subspace with fixed excitation quanta above the ground state. 
Then, in Sec.III, we examine a semiclassical description which does
not rely on these subspaces and can thus be extended to more
complicated Hamiltonians. 
With coherent state path integrals the leading order of the
propagator is determined by solving the Euler-Lagrange equations for the
classical path. In Sec.IV we consider contributions from fluctuations 
about the dominant path and show that they lead to a decay of the
excited two-level system by  spontaneous emission.

\section{The Jaynes-Cummings model}
The Jaynes-Cummings model is characterized by the Hamiltonian 
\begin{equation}
H= a^{\dagger} a +(1+\Delta) S_z + \lambda (a S_{+}+a^{\dagger}S_{-}),
\label{jc1}
\end{equation}
where $a$ is the canonical annihilation operator of a bosonic field
mode with frequency $\omega$ and $S_{\pm}=S_x\pm iS_y$, $S_z$ are 
operators of a spin-$\frac{1}{2}$ describing two levels of an 
 atomic system with energy difference $\hbar \omega_o$. 
There are two dimensionless parameters, the detuning
$\Delta=(\omega_o-\omega)/\omega$ and the  coupling
strength $\lambda=g/\omega$. We use units with $\omega=1$ and
$\hbar=1$.  It is well known that the Jaynes-Cummings model 
allows apart from $H$ for another time independent operator 
\cite{ackerhalt}
\begin{equation}
N= a^{\dagger} a + S_z,
\label{jc2}
\end{equation}
which measures the number of excitation quanta in the system.
Hence, the time evolution operator is of the  form 
\begin{equation}
U(T)=e^{-i H T }= e^{-i N T} e^{-i C T},
\label{jc3}
\end{equation}
where $C=H-N$. Representing the spin operators in the eigenbasis of
$S_z$ formed by the eigenvectors $|\uparrow\rangle$ and 
$|\downarrow\rangle$, the first factor in Eq.(\ref{jc3}) may be
written as  
\begin{equation}
e^{-i N T} = e^{-i a^{\dagger}a\,T} 
\Bigl(
e^{-\frac{i}{2}T} |\uparrow\rangle\langle\uparrow|
+e^{+\frac{i}{2}T}|\downarrow\rangle\langle\downarrow| 
\Bigr).
\label{jc4}
\end{equation}
Introducing further the eigenkets of $a^{\dagger}a$,
 invariant subspaces are distinguished. In particular
the kets 
$|\uparrow n -1\rangle\equiv |\uparrow\rangle|n-1\rangle$ and 
$|\downarrow n\rangle\equiv |\downarrow\rangle|n\rangle$
span the subspace with  $N=(n-\frac{1}{2})$. In this subspace
 the time independent operator $C$ generates $SU(2)$
dynamics. This can be seen explicitly by introducing the
operators 
\begin{eqnarray}
J_x&=&\frac{1}{2}\Bigl( |\uparrow n-1\rangle\langle\downarrow n|
		       +|\downarrow n\rangle\langle\uparrow n-1|\Bigr)
\nonumber\\
J_y&=&\frac{i}{2}\Bigl(-|\uparrow n-1\rangle\langle\downarrow n|
		       +|\downarrow n\rangle\langle\uparrow n-1|\Bigr)
\nonumber\\
J_z&=&\frac{1}{2}\Bigl( |\uparrow n-1\rangle\langle\uparrow n-1|
		       -|\downarrow n\rangle\langle\downarrow n|\Bigr),
\label{jc5}
\end{eqnarray}
describing the angular momentum of a spin-$\frac{1}{2}$. 
In terms of these  spin operators we have
\begin{equation}
C= 2\lambda \sqrt{n}J_x+ \Delta J_z,
\label{jc6}
\end{equation}
and we see that in this subspace $C$ gives indeed rise to pure $SU(2)$
dynamics. Accordingly, the propagator may 
be worked out exactly by a semiclassical approach  with
path integrals in the spin coherent state representation
\begin{equation}
|\vartheta\,\varphi\rangle=
e^{-i\varphi J_z} e^{-i\vartheta J_y}
|\uparrow n-1\rangle.
\label{jc7}
\end{equation}
Following the lines of \cite{alscher}, we write the spin coherent propagator
as a regularized path integral
\begin{equation}
\langle \vartheta''\,\varphi'' | e^{-i C T} | \vartheta'\,\varphi'\rangle 
 = \lim_{\nu \to\infty} \int d\mu
   \exp\left\{i\, S[\vartheta(t),\varphi(t)]\right\},
\label{jc8}
\end{equation}
with the action 
\begin{equation}
S[\vartheta(t),\varphi(t)]=
\int_{0}^{T} dt \biggl[\frac{1}{2}\cos(\vartheta)\dot\varphi 
-C(\vartheta,\varphi)\biggr].
\label{jc9}
\end{equation}
Here the operator $C$ is represented as 
\begin{eqnarray}
C(\vartheta,\varphi)&=& 
\langle \vartheta\,\varphi | C |\vartheta\,\varphi\rangle.
\nonumber\\
&=& \lambda \sqrt{n} \sin(\vartheta)\cos(\varphi)
+\frac{\Delta}{2}\cos(\vartheta).
\label{jc10}
\end{eqnarray}
The spherical  Wiener measure \cite{klauder,daubechies}
\begin{equation}
d\mu =M\prod_{t=0}^{T} d\cos(\vartheta(t))d\varphi(t) \,
\exp{ \biggl\{
- \frac{1}{4  \nu  }\int_{0}^{T} dt\, \Bigl[ 
\dot\vartheta^2+ \sin^2(\vartheta)\dot \varphi^2  \Bigr]\biggr\} },
\label{jc11}
\end{equation}
enforces continuous Brownian motion paths on the sphere 
($M$ is a normalization factor).
This measure gives rise to a regularization dependent action 
\begin{eqnarray}
S_{\nu}[\vartheta(t),\varphi(t)] 
&=& \int_{0}^{T} d t \biggl\{ \frac{i }{4\nu} 
\left[ \dot\vartheta^2 + \sin^2(\vartheta)\dot \varphi^2 \right] 
+\frac{1}{2}\cos(\vartheta)\dot\varphi 
- C(\vartheta,\varphi) \biggr\}.
\label{jc12}
\end{eqnarray}
Now, in the semiclassical expansion,  we separate the paths 
$\cos(\vartheta) =\cos(\vartheta_{cl})+ x/\sqrt{s}$ and 
$\varphi=\varphi_{cl}+y/\sqrt{s}$ in
their classical parts and fluctuations around them.
The formal limit of large spin $s\rightarrow\infty$ expresses the classical
limit. To lowest order, in the Dominant Path Approximation (DOPA), the
semiclassical expansion gives
\begin{equation}
e^{i S_{cl}}=\exp\biggr\{-i\int_{0}^{T} dt \,
C(\bar \vartheta''(t),\bar\varphi''(t)) \biggr\}
\langle \vartheta''\,\varphi'' |  \vartheta'\,\varphi'\rangle,
\label{jc13}
\end{equation}
where $(\bar\vartheta',\bar\varphi')$ and
$(\bar\vartheta'',\bar\varphi'')$, respectively, 
describe the starting point and  endpoint of the classical
trajectory $(\bar\vartheta(t),\bar\varphi(t))$, $0\le t \le T$.
For convenience let us introduce the complex variables 
\begin{eqnarray}
\zeta &=&\tan\left(\frac{\bar\vartheta}{2}\right)e^{i\bar\varphi} 
\nonumber\\
\eta &=&\tan\left(\frac{\bar\vartheta}{2}\right)e^{-i\bar\varphi}.
\label{jc14}
\end{eqnarray}
Then, the dominant path is determined by
\begin{eqnarray} 
\dot\zeta &=&-i\lambda\sqrt{n}(1-\zeta^2) +i\Delta\zeta
\nonumber\\
\dot\eta &=&i\lambda\sqrt{n}(1-\eta^2) -i \Delta\eta,
\label{jc15}
\end{eqnarray}
with boundary conditions $\zeta(0)=\zeta'$ and $\eta(T)=\eta''$.
Hence, the endpoint of the classical trajectory obeys
\begin{eqnarray}
\zeta(T) &=&\frac{2\Omega_n\zeta'\cos(\Omega_n T)
+i\left[\Delta\zeta' -\lambda\sqrt{n}\right]\sin(\Omega_n T)}
{2\Omega_n\zeta'\cos(\Omega_n T)
-i\left[\lambda\sqrt{n}\,\zeta'+\Delta\right] \sin(\Omega_n T)}
\nonumber\\
\eta(T) &=&\eta'',
\label{jc16}
\end{eqnarray}
with the Rabi frequency 
\begin{equation}
\Omega_{n}=\sqrt{\lambda^2 n + \frac{\Delta^2}{4} }.
\label{jc17}
\end{equation}
In terms of the complex variables (\ref{jc14}) we get 
\begin{eqnarray}
C(\zeta(T),\eta'')&=&
\lambda\sqrt{n}\frac{\zeta(T)  + \eta}{1+\zeta(T)\eta''}+
\frac{\Delta}{2}\frac{1-\zeta(T)\eta''}{1+\zeta(T)\eta''}
\label{jc18}\\
&=&i\frac{d}{dT}\log\biggl\{(1+\zeta'\eta'')\cos(\Omega_n T )
\nonumber\\
&&\quad\quad\quad\quad
- \frac{i}{\Omega_n}\Bigl[
\lambda\sqrt{n}(\zeta'+\eta'')
+\frac{\Delta}{2}(1-\zeta'\eta'') \Bigr]\sin(\Omega_n T)  \biggr\}.
\nonumber
\end{eqnarray}
Now the integral in Eq.(\ref{jc13}) is readily solved and the
propagator in the DOPA takes the form 
\begin{eqnarray}
e^{i S_{cl}}
&=&a(T)\cos\left(\frac{\vartheta''}{2}\right)
\cos\left(\frac{\vartheta'}{2}\right)
e^{\frac{i}{2}(\varphi''-\varphi')}
+ b(T)\cos\left(\frac{\vartheta''}{2}\right)
\sin\left(\frac{\vartheta'}{2}\right)
e^{\frac{i}{2}(\varphi''+\varphi')}
\nonumber\\
&&- b^{*}(T)\sin\left(\frac{\vartheta''}{2}\right)
\cos\left(\frac{\vartheta'}{2}\right)
e^{-\frac{i}{2}(\varphi''+\varphi')}
+ a^{*}(T)\sin\left(\frac{\vartheta''}{2}\right)
\sin\left(\frac{\vartheta'}{2}\right)
e^{-\frac{i}{2}(\varphi''-\varphi')},
\label{jc19}
\end{eqnarray}
where 
\begin{eqnarray}
a(T)&=&\cos(\Omega_{n}T)-i\frac{\Delta}{2\Omega_{n}}\sin(\Omega_{n}T)
\nonumber\\
b(T)&=&-i\frac{\lambda\sqrt{n}}{\Omega_{n}}\sin(\Omega_{n}T)
\label{jc20}.
\end{eqnarray}
As discussed elsewhere \cite{alscher} for pure $SU(2)$ dynamics the
DOPA is exact and Eqs.(\ref{jc19}),(\ref{jc20}) give indeed the
exact propagator \cite{shoreknight}.

In more general situations, such as for the case without rotating wave
approximation \cite{belobrov,milonni},  the system cannot be separated into 
invariant subspaces. Therefore it would be interesting to consider a
semiclassical expansion that does not rely on the $SU(2)$ generators
(\ref{jc5}).

\section{Semiclassical dynamics with coherent state path integrals}
In order to formulate a general semiclassical theory for a coupled
spin boson problem we make use of product coherent states
\begin{equation}
|\vartheta\,\varphi\,p\,q\rangle =e^{-i\varphi S_z}e^{-i\vartheta
S_y}e^{i(pQ-qP)}|\uparrow\rangle |0\rangle.
\label{jc21}
\end{equation}
These states are generated by momentum and space translations  
of the normalized vacuum state $|0\rangle$ and $SU(2)$ rotations of the
$S_z$ eigenstate $|\uparrow\rangle$. Again, the semiclassical
approximation is based on  the coherent state path integral 
representation. We express the propagator as 
\begin{equation}
\langle \vartheta''\,\varphi''\,p''\,q'' | U(t) 
| \vartheta'\,\varphi'\,p'\,q'\rangle = \lim_{\nu_a,\nu_b \to\infty} \int d
\mu_{a} d\mu_{b}
   \exp\left\{i\, S[p(t),q(t),\vartheta(t),\varphi(t)]\right\},
\label{jc22}
\end{equation}
with the action
\begin{equation}
S[p(t),q(t),\vartheta(t),\varphi(t)]= 
\int_{0}^{T} d t \biggl[
\frac{1}{2}(p \dot q -\dot p q)+
\frac{1}{2}\cos(\vartheta)\dot\varphi - 
H(\vartheta,\varphi,p,q) \biggr]. 
\label{jc23}
\end{equation}
For the Jaynes-Cummings model the Hamiltonian takes the form 
\begin{eqnarray}
H(\vartheta,\varphi,p,q)   &=&
\langle \vartheta\,\varphi\,p\,q | H | 
\vartheta\,\varphi\,p\,q \rangle 
\label{jc24}\\
&=&
\frac{1}{2}(p^2+q^2)
+\frac{1+\Delta}{2}\cos(\vartheta) 
+\frac{\lambda}{2\sqrt{2}}\left[
\sin(\vartheta)e^{i\varphi}(q+ip)+
\sin(\vartheta)e^{-i\varphi}(q-ip)\right].
\nonumber
\end{eqnarray}
Here, the  canonical coherent state path integral
is regularized by the flat Wiener measure \cite{klauder,daubechies}
\begin{equation}
d\mu_a =M_a\prod_{t=0}^{T} d p(t) d q(t)  \,\exp{ \biggl\{
- \frac{1}{2\nu_a}\int_{0}^{T} dt 
\Bigl[\dot q^2+ \dot p^2 \Bigr]\biggr\}},
\label{jc25}
\end{equation}
while the spin paths are again regularized by the spherical
Wiener measure 
\begin{equation}
d\mu_b=M_b\prod_{t=0}^{T} d\cos(\vartheta(t))d\varphi(t) \,
\exp{ \biggl\{
- \frac{1}{4  \nu_b}\int_{0}^{T} dt \Bigl[ 
\dot\vartheta^2+ \sin^2(\vartheta)\dot \varphi^2  \Bigr]\biggr\} }.
\label{jc26}
\end{equation}
These measures give rise to the regularization dependent action 
\begin{eqnarray}
S_{\nu_a,\nu_b}[\vartheta(t),\varphi(t),p(t),q(t)]&=& 
\int_{0}^{T} d t \biggl\{
\frac{i}{2\nu_a}\left[\dot q^2+ \dot p^2 \right]+
\frac{i }{4\nu_b} 
\left[\dot\vartheta^2 + \sin^2(\vartheta)\dot \varphi^2 \right] 
\nonumber\\
&&
\quad\quad +
\frac{1}{2}(p \dot q -\dot p q)+
\frac{1}{2}\cos(\vartheta)\dot\varphi - 
H(\vartheta,\varphi,p,q) \biggr\}.
\label{jc27}
\end{eqnarray}
In the semiclassical expansion we split the paths 
$p=p_{cl}+x_a$, $q=q_{cl}+y_a$, 
$\cos(\vartheta)=\cos(\vartheta_{cl}) + x_b / \sqrt{s}$ and
$\varphi=\varphi_{cl} + y_b / \sqrt{s}$ 
in their classical parts and fluctuations around them. 
Restricting ourselves to the DOPA, we obtain the propagator
\begin{eqnarray}
e^{i S_{cl} }&=&
\sqrt{\frac{\sin(\vartheta')\sin(\vartheta'')}
{\sin(\bar\vartheta')\sin(\bar\vartheta'')} }
\exp\biggl\{-\frac{1}{2}\Bigl[q'' \bar{p}''-\bar{q}'' p'' 
+\bar{q}'p' -q'\bar{p}' \Bigr]\biggr\} 
\nonumber\\ 
&&\times 
\exp\Biggl\{ i\int_{0}^{T} d t \biggl[ 
\frac{1}{2}\cos(\bar\vartheta)\dot{\bar{\varphi}}
+\frac{1}{2}(\bar p \dot{\bar q} -\dot{\bar p} \bar q)
- H(\bar \vartheta,\bar\varphi,\bar p,\bar q) \biggr]  \Biggr\}.
\label{jc28}
\end{eqnarray}
While for $\lambda=0$ this approximation yields the exact propagator, 
this property is lost for the interacting system.
Introducing the complex variables \cite{weissman,perelomov2}
\begin{eqnarray}
\alpha&=&\frac{1}{\sqrt{2}}(\bar q+i \bar p) 
\nonumber\\
\beta&=&\frac{1}{\sqrt{2}}(\bar q-i \bar p)
\nonumber\\
\zeta &=&\tan\left(\frac{\bar\vartheta}{2}\right)e^{i\bar\varphi} 
\nonumber\\
\eta &=&\tan\left(\frac{\bar\vartheta}{2}\right)e^{-i\bar\varphi},
\label{jc29}
\end{eqnarray}
Eq.(\ref{jc28}) may be expressed as 
\begin{eqnarray}
e^{i S_{cl}}&=&
 \exp\left\{-\frac{1}{2}\Bigl[ 
|\alpha'|^2+|\beta''|^2 -\alpha(T)\beta''-\alpha'\beta(0)
\Bigr]\right\} 
\sqrt{\frac{(1+\zeta'\eta(0))(1+\zeta(T)\eta'')}
{(1+\zeta'\eta')(1+\zeta''\eta'')} }
\nonumber\\ 
&&\times
\left(\frac{\zeta'' \eta'}{\zeta' \eta''}\right)^\frac{1}{4}
\exp\Biggl\{ i\int_{0}^{T} d t \biggl[ 
\frac{i}{2}\left( \dot\alpha \beta -\alpha\dot\beta \right)
+\frac{i}{2}\frac{\dot\zeta \eta-\zeta\dot\eta}{1+\zeta\eta}  
- H\left(\alpha,\beta,\zeta,\eta\right)\biggr]
 \Biggr\}, 
\label{jc30}
\end{eqnarray}
where the Hamiltonian (\ref{jc24}) reads
\begin{equation}
H(\alpha,\beta,\zeta,\eta)=\alpha\beta +
\frac{1+\Delta}{2}\frac{1-\zeta\eta}{1+\zeta\eta}+
\lambda\frac{\alpha\zeta  + \beta\eta}{1+\zeta\eta}.
\label{jc31}
\end{equation}
Now, the DOPA propagator is determined by the 
dominant path obeying the classical equations of motion 
\begin{eqnarray}
\dot \alpha&=&-i\Bigl[\alpha
+\lambda\frac{\eta}{1+\zeta\eta}\Bigr]
\nonumber\\
\dot \beta&=&i\Bigl[\beta
+\lambda\frac{\zeta}{1+\zeta\eta}\Bigr]
\nonumber\\
\dot \zeta &=&i\Bigl[(1+\Delta)\zeta -\lambda(\beta-\alpha\zeta^2)\Bigr]
\nonumber\\
\dot \eta &=&-i\Bigl[(1+\Delta)\eta -\lambda(\alpha-\beta\eta^2)\Bigr],
\label{jc33}
\end{eqnarray}
with the boundary conditions 
\begin{eqnarray}
\alpha(0)&=&\frac{1}{\sqrt{2}}(q'+i\,p') \nonumber\\
\beta(T)&=&\frac{1}{\sqrt{2}}(q''-i\,p'')\nonumber\\
\zeta(0) &=&\tan\left(\frac{\vartheta'}{2}\right)e^{i\varphi'} \nonumber\\
\eta(T) &=&\tan\left(\frac{\vartheta''}{2}\right)e^{-i\varphi''}.
\label{jc34}
\end{eqnarray}
The system of differential equations (\ref{jc33}) gives rise to a
Hamiltonian vector field. Extending  results in
\cite{perelomov2,provost}, one sees that  this Hamiltonian dynamics is
identical to the classical mechanics of a spin on the two-sphere
coupled to the phase space degrees of freedom of a one dimensional
harmonic oscillator.  Since the covariant divergence of the
Hamiltonian vector field vanishes, this dynamical system is
conservative and no attractor can occur.
The coupled differential equations (\ref{jc33}) with
conditions (\ref{jc34}) express a nonlinear boundary value
problem. We can find a solution exploiting the invariance of the
action (\ref{jc23}) under phase transformations 
\begin{eqnarray}
\zeta &\rightarrow & \zeta e^{i\Lambda} \nonumber\\
\eta &\rightarrow &\eta e^{-i\Lambda} \nonumber\\
\alpha &\rightarrow &\alpha e^{-i\Lambda} \nonumber\\
\beta &\rightarrow &\beta e^{i\Lambda}.
\label{jc35}
\end{eqnarray}
The corresponding integral of motion is
\begin{equation}
N(\alpha,\beta,\zeta,\eta)= \alpha\beta + \frac{1}{2}\frac{1-\zeta\eta}{1+\zeta\eta}.
\label{jc36}
\end{equation}
Therefore the Hamiltonian dynamical system  becomes integrable
by the theorem of Liouville-Arnold \cite{perelomov1}. 
Particularly, by setting $u=(1-\zeta\eta)/(1+\zeta\eta)$, we reduce the
system  to a  one-dimensional problem of the form 
\begin{equation}
\frac{1}{2}{\dot u}^2 +V(u)=0,
\label{jc37}
\end{equation}
with the   cubic potential
\begin{equation}
V(u)=\lambda^2(u^3+a_2 u^2+a_1 u + a_0).
\label{jc38}
\end{equation}
The coefficients read
\begin{eqnarray}
a_0&=&-2N +2\frac{C^2}{\lambda^2}\nonumber\\
a_1&=&1+2\frac{\Delta C}{\lambda^2} \nonumber\\
a_2&=&2N+\frac{\Delta^2}{2\lambda^2},
\label{jc39}
\end{eqnarray}
where
\begin{equation}
C(\alpha,\beta,\zeta,\eta) = \lambda\frac{\alpha\zeta+\beta\eta}{1+\zeta\eta}
+\frac{\Delta}{2}\frac{1-\zeta\eta}{1+\zeta\eta}.
\label{jc40}
\end{equation}
Although the potential $V(u)$ is time independent, the boundary values
(\ref{jc34}) enforce the coefficients in Eq.(\ref{jc39}) to depend on
the end time $T$, and the form of $V(u)$ changes with  $T$. 
Next we set $v=u+a_2 /3$ and rewrite  Eq.(\ref{jc37}) as 
\begin{equation}
\frac{2}{\lambda^2}{\dot v}^2 = 4 v^3 - g_2 v - g_3.
\label{eq41}
\end{equation}
This is just the differential equation solved by the  Weierstrass elliptic
function  \cite{tricomi} $\wp (\frac{\lambda}{\sqrt{2}}t ; g_2 ; g_3)$
with the  invariants 
\begin{eqnarray}
g_2&=&-4 \left(a_1-\frac{1}{3}a_2^2 \right)\nonumber\\
g_3&=&-\frac{4}{3}\left(\frac{2}{9}a_2^2 - a_1\right)a_2 -4a_0.
\label{jc42}
\end{eqnarray}
In the following we suppress these invariants in the list of arguments
of the function $\wp$. Now, the solution  of Eq.(\ref{jc37}) becomes 
\begin{equation}
u(t)=-\frac{a_2}{3} + \wp\left(A_1 + \frac{\lambda}{\sqrt{2}}t \right),
\label{jc43}
\end{equation}
where 
\begin{equation}
A_1 = \wp^{-1}\left( \frac{a_2}{3} + \frac{1-\zeta'\eta(0)}{1+\zeta'\eta(0)}\right)
\label{jc44}
\end{equation}
is determined by the inverse Weierstrass function $\wp^{-1}$. 
Making use of the solution (\ref{jc43}), the equations of motion lead
to elliptic integrals which  can be solved in terms of the 
Weierstrass elliptic functions $\wp$, $\zeta_w$ and $\sigma_w$ 
\cite{tricomi}. After some algebra one finds  for the field coordinates 
\begin{eqnarray}
\alpha(t)&=&\alpha' 
\left[\frac{1}{2\alpha'\beta(0)}\frac{\sigma_w (A_2+A_1)}{\sigma_w (A_2-A_1)}\right]^{1/2} 
\nonumber\\
&& \times
\exp\Bigl\{\frac{\lambda}{\sqrt{2}}\zeta_w(A_2)\, t  \Bigr\}
\exp\Bigl\{-i\left(1+\frac{\Delta}{2}\right) t \Bigr\}
\nonumber\\
&& \times
\left[ 
\left(2N + \frac{a_2}{3} -\wp(A_1+ \frac{\lambda}{\sqrt{2}}t) \right)
\frac{\sigma_w (A_2-A_1+\frac{\lambda}{\sqrt{2}}t)}
{\sigma_w (A_2+A_1+\frac{\lambda}{\sqrt{2}}t)} \right]^{1/2},
\label{jc45}
\end{eqnarray}
and
\begin{eqnarray}
\beta(t)&=&\beta''
\left[
\frac{1}{2\alpha(T)\beta''}
\frac{\sigma_w(A_2 -A_1-\frac{\lambda}{\sqrt{2}}T)}
{\sigma_w( A_2 + A1+\frac{\lambda}{\sqrt{2}}T) } \right]^{1/2}
\nonumber\\
&& \times
\exp\Bigl\{\frac{\lambda}{\sqrt{2}}\zeta_w(A_2)(T-t) \Bigr\}
\exp\Bigl\{-i\left(1+\frac{\Delta}{2}\right) (T-t) \Bigr\}
\nonumber\\
&& \times
\left[\left(2N + \frac{a_2}{3} -\wp(A_1+ \frac{\lambda}{\sqrt{2}}t)\right)
\frac
{\sigma_w (A_2+A_1-\frac{\lambda}{\sqrt{2}}t)}
{\sigma_w (A_2-A_1-\frac{\lambda}{\sqrt{2}}t)} \right]^{1/2},
\label{jc46}
\end{eqnarray}
where
\begin{equation}
A_2= \wp^{-1}\left(\frac{a_2}{3} + 2N \right). 
\label{jc47}
\end{equation}
The spin variables are found to read
\begin{eqnarray}
\zeta(t) &=& \zeta'\left[
\frac{1}{\zeta'\eta(0)} \frac{\sigma_w(A_3+A_1)}{\sigma_w(A_3-A_1)}
\frac{\sigma_w(A_4+A_1)}{\sigma_w(A_4-A_1)} \right]^{1/2}
\nonumber\\
&& \times
\exp\Bigl\{\frac{\lambda}{\sqrt{2}} [\zeta_w(A_3)+\zeta_w(A_4)]\,t \Bigr\}
\exp\{ it\}
\nonumber\\
&& \times
\left[
\frac{ 1 + \frac{a_2}{3} -\wp(A_1+ \frac{\lambda}{\sqrt{2}}t)}
{1 - \frac{a_2}{3} + \wp(A_1+ \frac{\lambda}{\sqrt{2}}t)}
\frac{\sigma_w (A_3-A_1+\frac{\lambda}{\sqrt{2}}t)}
{\sigma_w (A_3+A_1+\frac{\lambda}{\sqrt{2}}t)} 
\frac{\sigma_w (A_4-A_1+\frac{\lambda}{\sqrt{2}}t)}
{\sigma_w (A_4+A_1+\frac{\lambda}{\sqrt{2}}t)} \right]^{1/2}
\label{jc49}
\end{eqnarray}
and 
\begin{eqnarray}
\eta(t) &=&\eta''\left[
\frac{1}{\zeta(T)\eta''} 
\frac{\sigma_w(A_3+A_1+\frac{\lambda}{\sqrt{2}}T)}
{\sigma_w(A_3-A_1+\frac{\lambda}{\sqrt{2}}T)}
\frac{\sigma_w(A_4+A_1+\frac{\lambda}{\sqrt{2}}T)}
{\sigma_w(A_4-A_1+\frac{\lambda}{\sqrt{2}}T)} \right]^{1/2}
\nonumber\\
&& \times
\exp\Bigl\{\frac{\lambda}{\sqrt{2}} 
[\zeta_w (A_3)+\zeta_w (A_4)](T-t) \Bigr\}\exp\{ i(T-t)\}
\nonumber\\
&& \times
\left[
\frac{ 1 + \frac{a_2}{3} - \wp(A_1+ \frac{\lambda t}{\sqrt{2}})}
{1 - \frac{a_2}{3} + \wp(A_1+ \frac{\lambda t}{\sqrt{2}})}
\frac{\sigma_w (A_3-A_1+\frac{\lambda}{\sqrt{2}}t)}
{\sigma_w (A_3+A_1+\frac{\lambda}{\sqrt{2}}t)} 
\frac{\sigma_w (A_4-A_1+\frac{\lambda}{\sqrt{2}}t)}
{\sigma_w (A_4+A_1+\frac{\lambda}{\sqrt{2}}t)} \right]^{1/2}
\label{jc50}
\end{eqnarray}
where
\begin{eqnarray}
A_3&=& \wp^{-1}\Bigl(\frac{a_2}{3} - 1\Bigr)
\nonumber\\
A_4&=& \wp^{-1}\Bigl(\frac{a_2}{3} + 1\Bigr).
\label{jc51}
\end{eqnarray}
The solutions (\ref{jc45})-(\ref{jc51}) give the dominant path in
terms of the known initial values $\alpha(0)=\alpha'$,
$\zeta(0)=\zeta'$ and final values $\beta(T)=\beta''$,
$\eta(T)=\eta''$ and as implicit functions of the unknown 
initial values $\beta(0)$, $\eta(0)$ and  final values $\alpha(T)$,
$\zeta(T)$.  
Two of these unknowns have to be determined numerically. For instance,
from Eqs.(\ref{jc45}) and (\ref{jc49}) we obtain two transzendental
equations for $\alpha(T)$  and $\zeta(T)$ that can be solved 
by a root search procedure.  Then, the two other unknowns  
can be found from the two constants $C$ and $N$.

Having determined the semiclassical trajectory, we may insert the
result into Eq.(\ref{jc30}) and determine the DOPA-Propagator.
Since this propagator obeys a semiclassical Schr\"odinger
equation [see Appendix A], an alternative representation of the
propagator reads
\begin{equation}
e^{i S_{cl}}
=\exp\biggl\{
-i \int_{0}^{T} dt\, H(\alpha(t),\beta'',\zeta(t),\eta'') \biggr\}
\langle\vartheta''\,\varphi''|\vartheta'\,\varphi'\rangle
\langle p''\,q''|p'\,q'\rangle.
\label{jc59}
\end{equation}
With this representation the DOPA propagator is just determined  by the  
the endpoint of the classical path.

Although the dynamical system (\ref{jc33}) is conservative, it gives rise to 
stationary states. These are the  fix points
$(\alpha,\beta,\zeta_N,\eta_N)=(0,0,0,0)$ and  
$(\alpha,\beta,\rho_S,\sigma_S)=(0,0,0,0)$, where $\rho=1/\zeta$ and
$\sigma=1/\eta$. These points correspond to the states 
$|\uparrow\,0 \rangle$ and $|\downarrow\,0\rangle$ referred to
as north pole and south pole, henceforth. For a linear stability
analysis we just have to linearize the spin terms 
since  the equations of motion (\ref{jc33}) are already linear in  the
oscillator variables. Expanding about $(\zeta_N,\eta_N)$ we find
\begin{equation}
\frac{d}{dt}
\left(\begin{array}{c}
\delta\zeta \\ \beta \\  \delta\eta \\ \alpha \\  
\end{array}\right)
=
i\left(\begin{array}{c c c c}
1+\Delta &-\lambda &0 &0   \\  
\lambda  &     1   &0 &0   \\
0&0 & -1-\Delta &\lambda   \\  
0&0 & -\lambda  &   -1      \\
\end{array}\right)
\left(\begin{array}{c}
\delta\zeta \\ \beta \\  \delta\eta \\\alpha \\  
\end{array}\right),
\label{jc52}
\end{equation}
and two  invariant subspaces  in the variables  
$(\delta\zeta,\beta)$ and $(\delta\eta,\alpha)$ appear.
The solution satisfying the boundary conditions (\ref{jc34}) becomes
\begin{eqnarray}
\alpha(t)&=&\frac{1}{\cosh(\Omega_N T)}
\left\{\alpha' e^{-i \omega_m t}\cosh\left[\Omega_N (T-t)\right]
-i\eta''e^{i\omega_m (T-t)}\sinh(\Omega_N t)\right\} 
\nonumber\\
\beta(t)&=&\frac{1}{\cosh(\Omega_N T)}
\left\{\beta''e^{-i\omega_m (T-t)}\cosh(\Omega_N t)
-i\zeta'e^{i\omega_m t}\sinh\left[\Omega_N (T-t)\right]\right\} 
\nonumber\\
\delta\zeta(t)&=&\frac{1}{\cosh(\Omega_N T)}
\left\{\zeta'e^{i\omega_m t}\cosh\left[\Omega_N (T-t)\right]
-i\beta''e^{-i\omega_m (T-t)}\sinh(\Omega_N t)\right\} 
\nonumber\\
\delta\eta(t)&=&\frac{1}{\cosh(\Omega_N T)}
\left\{\eta''e^{i\omega_m (T-t)}\cosh(\Omega_N t)
-i\alpha'e^{-i \omega_m t}\sinh\left[\Omega_N (T-t)\right]\right\}
\label{jc53}
\end{eqnarray}
with the frequencies
\begin{eqnarray}
\omega_m&=& 1+\frac{\Delta}{2} \nonumber\\
\Omega_N&=&\sqrt{\lambda^2- \frac{\Delta^2}{4} }.
\label{eq54}
\end{eqnarray}
Note that for long times the dominant path converges to the 
corresponding boundary value and no 
oscillations around the north pole take place anymore.

In the same way, we linearize the motion around the south pole. Now 
invariant subspaces appear in the variables  
($\delta\rho,\alpha$) and ($\delta\sigma,\beta$)  
\begin{equation}
\frac{d}{dt}
\left(\begin{array}{c}
\delta\rho \\ \alpha \\  \delta\sigma \\ \beta \\  
\end{array}\right)
=
i\left(\begin{array}{c c c c}
-1-\Delta &-\lambda & 0 & 0   \\  
-\lambda  &     -1   & 0 & 0 \\
0 & 0 & 1+\Delta &\lambda   \\  
0 & 0 & \lambda  &   1     \\
\end{array}\right)
\left(\begin{array}{c}
\delta\rho \\ \alpha \\  \delta\sigma \\\beta \\  
\end{array}\right),
\label{jc55}
\end{equation}
with the  solution 
\begin{eqnarray}
\alpha(t)&=&\alpha'e^{-i\omega_m t}\cos(\Omega_S t)-
i\frac{1}{\zeta'}e^{-i \omega_m t}\sin(\Omega_S t)
\nonumber\\
\beta(t)&=& \beta''e^{-i\omega_m (T-t)}\cos\left[\Omega_S (T-t)\right] 
-i\frac{1}{\eta''}e^{-i\omega_m (T-t)}\sin\left[\Omega_S (T-t)\right] 
\nonumber\\
\delta\rho(t) &=&\frac{1}{\zeta'}e^{-i\omega_m t}\cos(\Omega_S t)-
i \alpha'e^{-i \omega_m t}\sin(\Omega_S t) 
\nonumber\\
\delta\sigma(t)&=& \frac{1}{\eta''}e^{-i\omega_m (T-t)}\cos\left[\Omega_S (T-t)\right] 
-i\beta''e^{-i\omega_m (T-t)}\sin\left[\Omega_S (T-t)\right] ,
\label{jc56}
\end{eqnarray}
where
\begin{equation}
\Omega_S=\sqrt{\lambda^2 + \frac{\Delta^2}{4}}.
\label{jc57}
\end{equation}
Here, the dominant path does not converge for long times but 
keeps on oscillating around the south pole. North pole and south pole
correspond to the local extrema of the cubic potential (\ref{jc38})
generated by the coupling of the spin-$\frac{1}{2}$ to a vacuum
field. Whenever the field becomes filled with bosons, these fix points
bifurcate into limit cycles. 

The presence of stationary states leads to strong deviations of the
DOPA propagator from the exact result for times large compared to
$\omega_o^{-1}$. In fact, for long times the semiclassical trajectory
approaches the saddle point of the cubic potential  and
stays there for most of the time. For the full quantum problem the
state $|\uparrow 0\rangle$ is not a steady state, rather it will decay
by spontaneous emission. In the semiclassical approximation
spontaneous emission arises from fluctuations about the classical path
that are neglected in the DOPA. Hence, to obtain useful results also
for long times, fluctuations about the north pole need to be taken into
account.

\section{Fluctuations}
The semiclassical expansion of the path integral (\ref{jc8}) 
leads to second order contributions in terms of Gaussian  fluctuation
path integrals.  Denoting by $(x_a,y_a)$ and $(x_b,y_b)$
deviations from the dominant path variables 
$(p,q)$ and $(\cos(\vartheta),\varphi)$, 
the semiclassical approximation takes the form 
\begin{equation}
\langle\vartheta''\,\varphi''\,p''\,q'' | U(T) | 
\vartheta'\,\varphi'\,p'\,q'\rangle_{sc} 
=e^{iS_{cl}}
\lim_{\nu_a,\nu_b\to\infty}
\int d \mu_a d \mu_b 
\exp\left\{i\,\delta^2 S[x_a(t),y_a(t),x_b(t),y_b(t)]\right\},
\label{jc60}
\end{equation}
with the boundary conditions $x_a(0)=x_a(T)=0$, $y_a(0)=y_a(T)=0$,
$x_b(0)=x_b(T)=0$, $y_b(0)=y_b(T)=0$. 
Since the  canonical Wiener measure (\ref{jc25}) is of quadratic form, the
measure of the fluctuation path integral becomes
\begin{equation}
d\mu_{a} =\prod_{t=0}^{T} \frac{1}{2\pi}d x_a (t) d y_a (t) \,
\exp{ \biggl\{
- \frac{1}{2\nu_a}\int_{0}^{T} dt 
\Bigl[\dot{x}_{a}^{2}+ \dot{y}_{a}^{2} \Bigr]\biggr\}},
\label{jc61}
\end{equation}
which is of the same form as the original coherent state path measure. 
On the other hand, the  spin measure (\ref{jc26}) is not
quadratic, and the dominant path $(\vartheta(t),\varphi(t))$ 
cannot be separated from the fluctuation variables $x_b$ and $y_b$. We
have 
\begin{eqnarray}
d\mu_b&=&
\prod_{t=0}^{T} \frac{2s+1}{4\pi s}d x_b (t) d y_b (t) 
\exp \Biggl\{ - \frac{1}{2\nu_b}\int_{0}^{T} dt 
\biggl[ \frac{\dot x_b^2}{\sin^2(\vartheta)} + 
\sin^2(\vartheta)\,\dot y_b^2 
\nonumber\\
&& \quad\quad\quad\quad  
-\cos(\vartheta)\dot\varphi\,x_b\dot y_b 
+2\frac{\cos(\vartheta)\dot\vartheta}
   {\sin^3(\vartheta)}\, \dot x_b x_b 
+\Bigl( \dot\varphi^2 -\frac{ 2\cos^2(\vartheta)\dot\vartheta^2
   +\sin(2\vartheta)\ddot\vartheta }
    {2\sin^4(\vartheta)}\Bigr)\,x_b^2 
\biggr]\Biggr\},
\label{jc62}
\end{eqnarray}
and the regularization of the fluctuation path integral becomes in
general time dependent. 
However, when the dominant spin path is strictly independent of
time, $(\vartheta(t),\varphi(t))=(\vartheta_o,\varphi_o)$, 
the measure (\ref{jc62}) simplifies considerably and we get    
\begin{eqnarray}
d\mu_b&=&\prod_{t=0}^{T} \frac{2s+1}{4\pi s}d x_b(t) d y_b(t)  
\,\exp\biggl\{ - \frac{1}{2\nu_b}\int_{0}^{T} dt 
\Bigl[ \frac{\dot x_b^2}{\sin^2(\vartheta_o)} 
+\sin^2(\vartheta_o)\dot y_b^2 \Bigr]\biggr\}.
\label{jc63}
\end{eqnarray}
Then, after a canonical transformation 
\begin{eqnarray}
\tilde x_b &=&\frac{x_b}{\sin(\vartheta_o)}
\nonumber\\
\tilde y_b &=&\sin(\vartheta_o) y_b,
\label{jc67}
\end{eqnarray}
the measure(\ref{jc63}) takes for large $s$ the form of the 
canonical measure (\ref{jc61}) 
\begin{equation}
d\mu_{b} =
\int \prod_{t=0}^{T} \frac{1}{2\pi}d \tilde x_b (t) d \tilde y_b (t) \,
\exp{ \biggl\{
- \frac{1}{2\nu_b}\int_{0}^{T} dt 
\Bigl[\dot{\tilde{x}}_b^{2}+ \dot{\tilde y}_b^{2} \Bigr]\biggr\}}.
\label{jc68}
\end{equation}
Both measures give rise to the regularization dependent 
second order variational action 
\begin{eqnarray}
&&\delta^2 S_{\nu_a,\nu_b}[x_a(t),y_a(t),\tilde x_b (t),\tilde y_b (t)]
=\int_{0}^{T} dt \Bigl[\frac{i}{2\nu_a}(\dot x_a^2+\dot y_a^2)
+\frac{i}{2\nu_b}(\dot{\tilde{x}}_b^2+\dot{\tilde{y}}_b^2)
\nonumber\\
&&\quad\quad+
\frac{1}{2}(x_a\dot y_a -\dot x_a y_a)+
\frac{1}{2}(\tilde x_b \dot{\tilde{y}}_b -\dot{\tilde{x}}_b \tilde y_b)
-H_o(x_a,y_a,\tilde x_b,\tilde y_b,t)
\Bigr],
\label{jc69}
\end{eqnarray}
where the Hamiltonian  $H_{o}(t)$ is determined by the 
second order contributions of the Hamiltonian $H$ 
expanded around the dominant path
\begin{eqnarray}
H_o(x_a,y_a,\tilde x_b,\tilde y_b,t) 
&=& a_1 (t) x_a^2  +a_2 (t)x_a y_a + a_3(t) y_a^2
\nonumber\\
&&
+b_1(t) \tilde x_b^2
+b_2(t)\tilde x_b \tilde y_b 
+b_3(t)\tilde y_b^2
\nonumber\\
&&
+c_1(t)x_a \tilde x_b
+c_2(t)x_a \tilde y_b 
+c_3(t)y_a \tilde x_b
+c_4(t)y_a \tilde y_b, 
\label{jc70}
\end{eqnarray}
with the coefficients
\begin{eqnarray}
&&
a_1 (t) = \frac{1}{2}\frac{\partial^2 H}{\partial \bar q^2},\quad
a_2 (t)=\frac{\partial^2 H}{\partial \bar q \partial\bar p},\quad
a_3 (t)=\frac{1}{2}\frac{\partial^2 H}{\partial \bar p^2}
\nonumber\\
&&
b_1 (t)=\frac{\sin^2(\vartheta_o)}{2s}\frac{\partial^2 H}
	{\partial \cos(\bar\vartheta)^2} ,\quad
b_2 (t)= \frac{1}{s}\frac{\partial^2 H}
{\partial \bar\varphi\partial\cos(\bar\vartheta)},\quad
b_3 (t)=\frac{1}{2s\,\sin^2(\vartheta_o)}\frac{\partial^2 H}
	{\partial \bar\varphi^2}
\nonumber\\
&&
c_1 (t)=\frac{\sin(\vartheta_o)}{\sqrt{s}}\frac{\partial^2 H}
	{\partial \bar p \partial\cos(\bar\vartheta)} ,\quad
c_2 (t)= \frac{1}{\sqrt{s}\sin(\vartheta_o)}\frac{\partial^2 H}
	{\partial \bar p \partial \bar\varphi},\quad
\nonumber\\
&&
c_3 (t)=\frac{\sin(\vartheta_o)}{\sqrt{s}}\frac{\partial^2 H}
	{\partial \bar q \partial \cos(\bar\vartheta)}, \quad
c_4 (t)=\frac{1}{\sqrt{s}\sin(\vartheta_o)}\frac{\partial^2 H}
	{\partial \bar q \partial \bar\varphi}.
\label{jc71}
\end{eqnarray}
For large $s$, starting and end points in the coherent state
fluctuation path integral (\ref{jc60})  parameterize  states  
$|x_a(0)\,y_a (0)\,x_b (0)\,y_b (0)\rangle$ and
$|x_a(T)\,y_a (T)\,x_b (T)\,y_b (T)\rangle$ which correspond
to product vacuum states. Propagators leading to stationary
saddle points $(\vartheta_o,\varphi_o)$ may be represented  now as 
\begin{equation}
\langle\vartheta_{o}\,\varphi_{o}\,p''\,q''
 |U(T)|\vartheta_{o}\,\varphi_{o}\,p'\,q'\rangle_{sc}  =
e^{iS_{cl}} \langle 0\,0| U_{o}(T) |0\,0\rangle,
\label{jc72}
\end{equation}
with the unitary time evolution operator 
\begin{equation}
U_{o}(T)= {\cal T}_{t}\exp\biggl\{
-i \int_{0}^{T} dt \, H_o(t) \biggr\},
\label{jc72b}
\end{equation}
determined by the quadratic Hamiltonian 
\begin{eqnarray}
H_o (t)&=&
a_1(t)\,(Q_a^2-\frac{1}{2})+a_2(t)\,(P_aQ_a+Q_aP_a)+a_3(t)\,(P_a^2-\frac{1}{2})
\nonumber\\
&&+
b_1(t)\,(Q_b^2-\frac{1}{2})+b_2(t)\,(P_bQ_b+Q_bP_b)+b_3(t)\,(P_b^2-\frac{1}{2})
\nonumber\\
&&+
c_1(t)\,P_aP_b + c_2(t)\,P_a Q_b +  c_3(t)\,Q_a P_b+c_4(t)\,Q_a Q_b,
\label{jc73}
\end{eqnarray}
describing two driven coupled oscillators.

As we have seen in the previous section, for the  Jaynes-Cummmings 
model  the  north pole $|\uparrow 0\rangle$ becomes a steady state in
the DOPA, and it is essential to take fluctuations about this state
into account.  Unfortunately, the description  of the 
spin degrees of freedom with spherical coordinates 
leads to coordinate singularities. Particularly, the azimuthal 
angle $\varphi$ is undefined at the poles of the two-sphere.
To calculate fluctuations about the north pole  accurately, we change
the coordinate system by a rotation. Since rotations are isometrical
canonical transformations, the spin path measure (\ref{jc26}) stays invariant  
but the kinematical term is not preserved. Instead a phase factor
appears which only vanishes if starting and endpoint of the spin
coordinates are identical. 

Within the DOPA, the probability amplitude
to remain at the north pole is just a phase factor 
\begin{equation}
\langle \uparrow 0 | U(T) | \uparrow 0\rangle_{DOPA} =
e^{i S_{cl}} = \exp\Bigl\{-i\frac{1+\Delta}{2} T\Bigr\},
\label{jc75}
\end{equation}
and the north pole becomes a steady state. Taking now Gaussian
fluctuations into account we have 
\begin{equation}
\langle \uparrow 0 | U(T) | \uparrow 0\rangle_{sc}  =
e^{iS_{cl}} \langle 0\,0|U_{o}(T)|0\,0\rangle,
\label{jc74}
\end{equation}
where the the vacuum amplitude is determined by the time independent
Hamiltonian  
\begin{equation}
H_o=\frac{1}{2}\,(P_a^2+Q_a^2-1)-
\frac{1+\Delta}{2}\,(P_b^2+Q_b^2-1)
+ \lambda(P_b Q_a+Q_b P_a).
\label{jc76}
\end{equation}
For convenience we represent the  operators  $Q_a,P_a$ and $Q_b,P_b$
by corresponding creation and annihilation operators
$a,a^\dagger$ and $b,b^\dagger$  
\begin{equation}
H_o= - 1 +(1+\frac{\Delta}{2})( a a^\dagger   -b^\dagger b)
	-\frac{\Delta}{2}( a a^\dagger   +b^\dagger b)
	-i\lambda(a b -a^\dagger b^\dagger).
\label{jc77}
\end{equation}
Since  $a a^\dagger  -b^\dagger b$ commutes with $H_o$,  
we rewrite the time evolution operator in the form
\begin{equation}
U_{o}(T)=
\exp\Bigl\{-i\frac{\Delta T}{2} \Bigr\}
\exp\Bigl\{-i(1+\frac{\Delta}{2}) a^\dagger a\, T\Bigr\}
\exp\Bigl\{i(1+\frac{\Delta}{2}) b^\dagger b\, T\Bigr\}\, U_1(T),
\label{jc78}
\end{equation}
with $U_1(T)=\exp(-i H_1 T)$ and 
\begin{equation}
H_1= - \Bigl[ \frac{\Delta}{2}(a a^\dagger+ b^\dagger b )
+i \lambda(a b -a^\dagger b^\dagger) \Bigr].
 \label{jc79}
\end{equation}
The  operators $a a^\dagger + b^\dagger b$, $a b$ and 
$a^\dagger b^\dagger$ span the three dimensional $su(1,1)$ Lie
algebra with  commutators 
\begin{eqnarray}
\left[ a a^\dagger+ b^\dagger b, a b \right] &=& -2 a b
\nonumber\\
\left[ a a^\dagger+ b^\dagger b, a^\dagger b^\dagger \right] &=& 2a^\dagger b^\dagger
\nonumber\\
\left[a b ,  a^\dagger b^\dagger\right] &=&  a a^\dagger+ b^\dagger b.
\label{jc80}
\end{eqnarray}
For this algebra there is a decomposition into one-dimensional
$SU(1,1)$ transformations which holds for the whole group, i.e. for
all times \cite{wey}. We start with the ansatz
\begin{equation}
U_{1}(T) =
\exp\{\mu(T) a^\dagger b^\dagger\}
\exp\{\nu(T) a b\}
\exp\{\xi(T)(a a^\dagger+ b^\dagger b)\},
\label{jc81}
\end{equation}
which results in  the vacuum amplitude 
\begin{equation}
 \langle 0\,0|U_{o}(T)|0\,0\rangle=
\exp\Bigl\{-i\frac{\Delta}{2} T + \xi(T) \Bigr\}.
\label{jc82}
\end{equation}
Then, requiring that $U_1(T)$ obeys the Schr\"odinger equation 
$d /dT  U_1(T) = -i H_1 U_1(T)$,  we get  
the  relation 
\begin{eqnarray}
i\frac{\Delta}{2}(a a^\dagger+ b^\dagger b )+
\lambda (a^\dagger b^\dagger -a b)
&=&
\dot\mu a^\dagger b^\dagger
+\dot\nu e^{\mu\,a^\dagger b^\dagger}ab
\,e^{-\mu\,a^\dagger b^\dagger}
\nonumber\\
&&+\dot\xi e^{\mu\,a^\dagger b^\dagger}
e^{\nu\,ab}(a a^\dagger+ b^\dagger b )e^{-\nu\,ab}
 e^{-\mu\,a^\dagger b^\dagger},
\label{jc83}
\end{eqnarray}
where we have made use of the Baker-Campbell-Hausdorff
formula. Further, the commutation relations (\ref{jc80}) imply 
\begin{eqnarray}
e^{\nu\,ab}(a a^\dagger+ b^\dagger b )e^{-\nu\,ab}&=&
a a^\dagger+ b^\dagger b + 2\nu ab
\nonumber\\
e^{\mu\,a^\dagger b^\dagger}(a a^\dagger+ b^\dagger b )e^{-\mu\,a^\dagger b^\dagger} &=&
a a^\dagger+ b^\dagger b -2\mu a^\dagger b^\dagger
\nonumber\\
e^{\mu\,a^\dagger b^\dagger}ab\,e^{-\mu\,a^\dagger b^\dagger}&=&
ab-\mu(a a^\dagger+ b^\dagger b ) +\mu^2 a^\dagger b^\dagger.
\label{jc84}
\end{eqnarray}
Now, Eq.(\ref{jc83}) determines the time rate of change of the functions   
$\mu$, $\nu$ and $\xi$ by the  linear equations  
\begin{equation}
\left(\begin{array}{c}
\lambda \\ \lambda \\ i \Delta \\
\end{array}\right)
=
\left(\begin{array}{c c c }
1 & \mu^2 & -2\mu (1-\mu\nu)   \\  
0 &     -1   & -2\nu   \\
0 & -2\mu & 2(1-2\mu\nu) \\  
\end{array}\right)
 \left(\begin{array}{c}
\dot\mu   \\ \dot\nu \\ \dot\xi\\
\end{array}\right),
\label{jc85}
\end{equation}
which are readily solved with the initial conditions 
$\mu(0)=0$, $\nu(0)=0$ and $\xi(0)=0$. 
In particular, we get for the  function $\xi(T)$ in Eq.(\ref{jc81}) 
\begin{equation}
\xi(T)=i \Delta T + \log\left[\cos(\Omega T) 
-i\frac{\Delta}{2}\sin(\Omega T)\right],
\label{jc86}
\end{equation}
with the Rabi frequency 
\begin{equation}
\Omega = \sqrt{\lambda^2+ \frac{\Delta^2}{4} }.
\label{jc87}
\end{equation}
Hence, the vacuum amplitude (\ref{jc82}) becomes 
\begin{equation}
\langle 0\,0|U_o(T)|0\,0\rangle=
\exp\Bigl\{i\frac{\Delta T}{2} \Bigr\} 
\left[\cos(\Omega T) -i\frac{\Delta}{2}\sin(\Omega T)\right].
\label{jc88}
\end{equation}
and the semiclassical propagator with fluctuations 
\begin{equation}
\langle\uparrow 0|U(T)|\uparrow 0 \rangle_{sc} =
e^{-\frac{i}{2}T} 
\left[\cos(\Omega T) -i\frac{\Delta}{2}\sin(\Omega T)\right]
\label{jc89}
\end{equation}
includes  spontaneous emission  leading  to an
instability of the north pole. Eq.(\ref{jc89}) gives the exact matrix
element of the propagator sandwiched between north pole states.

When the field is initially and finally not in the vacuum state, the 
semiclassical propagator (\ref{jc60}) is no longer characterized by a fix
point path. An evaluation of the fluctuations about the semiclassical
path would then require numerical methods beyond the scope of this
article.

\acknowledgments
The authors would like to thank Joachim Ankerhold and  
J\"urgen Stockburger for valuable discussions.  
This work was supported by the Deutsche For\-schungs\-ge\-mein\-schaft
(Bonn)  through the Schwer\-punkt\-pro\-gramm 
``Zeit\-ab\-h\"angi\-ge Ph\"anomene und Methoden in Quan\-ten\-sys\-te\-men 
der Phy\-sik und Che\-mie''. 

\appendix
\section{Semiclassical Schr\"odinger equation}
Here we derive the semiclassical Schr\"odinger equation for the DOPA
propagator given in Eq.(\ref{jc30}). 
The time rate of change is readily evaluated, and 
after an integration by parts it may be  expressed as
\begin{eqnarray} 
\frac{\partial}{\partial T} e^{i S_{cl}}&=&
\frac{1}{2}\Biggl\{  -
\frac{\partial \alpha(T,T)}{\partial T}\beta''-
\alpha'\frac{\partial \beta(0,T)}{\partial T}-
\frac{\zeta'\frac{\partial \eta(0,T)}{\partial T}}{1+\zeta'\eta(0,T)}-
\frac{\frac{\partial \zeta(T,T)}{\partial T}\eta''}{1+\zeta(T,T)\eta''}
  \Biggr.
\nonumber\\
&& -
\left.\frac{\partial \alpha(t,T)}{\partial t}\right|_T  \beta'' +
\alpha(T,T) \left.\frac{\partial \beta(t,T)}{\partial t}\right|_T -
\frac{
\left.\frac{\partial\zeta(t,T)}{\partial t}\right|_T \eta'' -
\zeta(T,T)\left.\frac{\partial \eta(t,T)}{\partial t}\right|_T
} {1+\zeta(T,T)\eta''}
\nonumber\\
&& 
-2 i H(\alpha(T,T),\beta'',\zeta(T,T),\eta'') 
\nonumber\\
&&
-\Biggl[
\frac{\partial \alpha(t,T)}{\partial T}\beta(t,T)-
\alpha(t,T)\frac{\partial \beta(t,T)}{\partial T}+
\frac{
\frac{\partial\zeta(t,T)}{\partial T} \eta(t,T) -
\zeta(t,T)\frac{\partial \eta(t,T)}{\partial T}} 
{1+\zeta(T,T)\eta''}
\Biggr]_{t=0}^{t=T}
\nonumber\\
&&
-\int_{0}^{T} dt 
\Biggl[
\frac{\alpha(t,T)}{\partial T}
\left(   
\frac{\partial\beta(t,T)}{\partial t}-i\frac{\partial H}{\partial\alpha}    
\right)-
\frac{\beta(t,T)}{\partial T}
\left(
\frac{\partial\alpha(t,T)}{\partial t}+i\frac{\partial H}{\partial\beta}
\right) 
\Biggr.
\nonumber\\
&&\quad\quad
\Biggl.
\frac{\zeta(t,T)}{\partial T}
\left(
\frac{\frac{\partial\eta(t,T)}{\partial t}}{(1+\zeta(t,T)\eta(t,T))^2}
-i\frac{\partial H}{\partial\zeta}   
\right)  
\nonumber\\
&&\quad\quad
-\frac{\eta(t,T)}{\partial T}
\left(
\frac{\frac{\partial\zeta(t,T)}{\partial t}}{(1+\zeta(t,T)\eta(t,T))^2}
+i\frac{\partial H}{\partial\eta}   
\right)
 \Biggr]
\Biggr\} e^{i S_{cl}}.
\label{a53}\\
\end{eqnarray}
Using the classical equations of motions, the integral is found to
vanish. Then, we rewrite the remaining parts in the form
\begin{eqnarray} 
\frac{\partial }{\partial T}e^{i S_{cl}}&=& 
-i H(\alpha(T,T),\beta'',\zeta(T,T),\eta'')\nonumber\\
&&-\frac{1}{2}\left[
\beta''\left(
-\frac{\partial \alpha(T,T)}{\partial T}
+\left.\frac{\partial\alpha(t,T)}{\partial t}\right|_T
+\left.\frac{\partial\alpha(t,T)}{\partial T}\right|_T \right)
\right.
\nonumber\\
&&
+\alpha'\left(
-\frac{\partial \beta(0,T)}{\partial T}+
\left.\frac{\partial \beta(t,T)}{\partial T}\right|_0\right)
\nonumber\\
&&
+\alpha(T,T)\left(
-\left.\frac{\partial \beta(t,T)}{\partial t}\right|_T
-\left.\frac{\partial \beta(t,T)}{\partial T}\right|_T\right)
-\beta(0,T)\left.\frac{\partial \alpha(t,T)}{\partial T}\right|_0
\nonumber\\
&&
+\frac{ \eta''\left(
-\frac{\partial \zeta(T,T)}{\partial T}
+\left.\frac{\partial\zeta(t,T)}{\partial t}\right|_T
+\left.\frac{\partial\zeta(t,T)}{\partial T}\right|_T  \right)
+\zeta(T,T)\left(
-\left.\frac{\partial \eta(t,T)}{\partial t}\right|_T
-\left.\frac{\partial \eta(t,T)}{\partial T}\right|_T \right)}
{1+\zeta(T,T)\eta''}
\nonumber\\
&&\left.
+\frac{\zeta'\left(
-\frac{\partial \eta(0,T)}{\partial T}+
\left.\frac{\partial \eta(t,T)}{\partial T}\right|_0 \right)
-\eta(0,T)\left.\frac{\partial \zeta(0,T)}{\partial T}\right|_0}
{1+\zeta'\eta(0,T)}
\right]e^{i S_{cl}}, 
\label{a54}
\end{eqnarray}
where  most of the terms on the right hand site vanish.  Finally we get 
\begin{equation}
\frac{\partial}{\partial T} e^{i S_{cl}} 
=-i H\left(\alpha(T),\beta'',\zeta(T),\eta''\right)e^{i S_{cl}}.
\label{a55}
\end{equation}
Note that the matrix element of the Hamiltonian at the endpoint of the
dominant path  $(\alpha(T),\beta'',\zeta(T),\eta'')$ generates the
time rate of change of the  DOPA propagator and not the matrix element
of the  final state $| \vartheta''\,\varphi''\,p''\,q''\rangle$. 
For a spin-$\frac{1}{2}$ coupled to a classical field 
this  Schr\"odinger equation generates  the exact quantum mechanics
\cite{alscher}.

\end{document}